\documentclass{wqcd03}                 
\usepackage{amsmath}
\usepackage{graphicx}

\confname{QCD@Work 2003 - International Workshop on QCD, Conversano, Italy,
14--18 June 2003}

\title{High density effective theory results for two and three massive flavors color-superconductivity}

\author{M.~Ruggieri\addressmark{a}\addressmark{b} }

\address[a]{Dipartimento di Fisica, Universit\`a di Bari, I-70124 Bari, Italia}
\address[b]{I.N.F.N., Sezione di Bari, I-70124 Bari, Italia}

\newcommand{\be}{\begin{equation}}

\newcommand{\ee}{\end{equation}}
\newcommand{\bea}{\begin{eqnarray}}
\newcommand{\eea}{\end{eqnarray}}

\begin{document}

\begin{abstract}
We describe some topics related to massive two and three flavors color superconductivity (CSC),
obtained in the framework of high density effective theory (HDET); moreover we present a {\em modified}
Nambu-Jona Lasinio model with running - as opposite to fixed - ultraviolet cutoff, discussing the troubles
that one encounters with the former when discussing QCD at high density.
\end{abstract}

\maketitle

\section{Introduction}
The existence of color-superconductivity (CSC) at very large densities and very low temperatures is a well estabilished consequence of QCD\cite{Rajagopal:2000wf}. The study of this state of quark and gluons matter is not only interesting for itself, but could be very important to achieve a greater knowledge of the physics of compact stars. Indeed, the conditions of high density-low temperature could be realized for example in the core of a neutron star.

While at asymptotic densities one can use perturbative methods, at lower densities (as for example the ones that could be realized in the core of a neutron star) one can use effective field theories as Nambu-Jona Lasinio (NJL) models. Moreover, the phase diagram of the possible phases of CSC with massless flavors is well known, while the situation with massive quarks is under investigation.

In this paper we address to study different aspects of CSC with massive flavors, employing a modified NJL model and working in the framework of HDET\cite{Hong:1998tn} (for a review see\cite{Nardulli:2002ma}). As a first point, we consider the three flavor case, known in literature as color-flavor locked (CFL) phase, with massless $u,d$ quarks and massive $s$ quark. We are concerned with:
\begin{itemize}
\item the role of the ultraviolet cutoff in the NJL models
\item the relevance of the effects due to the strange quark mass
\end{itemize}
Moreover, we briefly face to the following problems for the two massive flavors (2SC) phase:
\begin{itemize}
\item study of the gap equation
\item dynamical properties of the pseudo-Goldstone $\eta'$.
\end{itemize}

\section{A modified NJL model}
\label{NJL}
This section and the following one are based on our paper\cite{Casalbuoni:2003cs}.

When the NJL interaction is used for modelling QCD at vanishing temperature and density, one can fix  the UV cutoff $\Lambda$ such as to
get realistic quark constituent masses; in this way one also gets the value of the NJL coupling constant $G$ for the choosen scale of
energy $\Lambda$\cite{Klevansky:qe}. In such a case $\Lambda$ is thought of as fixed once for all. This gives no problems at zero density;
however, leads to difficulties when one tries to simulate QCD at finite chemical potential. In fact, in HDET one takes as relevant degrees
of freedom all the fermions with momenta in a shell around the Fermi surface. The thickness of the shell is measured by a cutoff $\delta$,
which is the cutoff for momenta measured from the Fermi surface. The cutoff $\delta$ satisfies the bounding condition
$\Delta\ll\delta\ll\mu$, and is related to the NJL cutoff by means of the relation $\Lambda=\delta+\mu$.

This relation is problematic when one is interested in the behavior of the theory for varying $\mu$. The constraint $\Lambda=\mu+\delta$
would force $\delta$ to vanish for increasing $\mu$, starting from $\mu<\Lambda$,which does not correspond to the asymptotic behaviour
$(\mu\rightarrow\infty)$ QC behaviour\cite{Son:1998uk}. The uncorrect behavior of $\Delta$ arises because the model is taken to be valid
only for momenta up to $\Lambda$ which forbids  to go to values of $\mu$ of the order or higher than $\Lambda$. Clearly this constitutes an
obstacle in physical situations where the typical chemical potential is about 400 or 500 MeV (e.g. in compact stellar objects) with a
$\delta$ of the order 150 or 200 MeV. In fact it turns out difficult, if not impossible, to explore higher values of $\mu$ for any
reasonable choice of $\Lambda$.

We make a proposal to overcome this situation\footnote{For the sake of discussion we consider here the ideal case of massless quarks, the
more realistic case of a massive strange quark will be considered in the following.}, suggesting the following procedure that allows the
introduction of a {\em running} NJL coupling constant. We write the Nambu-Jona Lasinio equations with a three dimensional cutoff $\Lambda$
\footnote{For a detailed account see Ref.\cite{Klevansky:qe} and Eqs.(1) and (2) of\cite{Casalbuoni:2003cs}.}. From the $f_\pi$ equation
(with $f_\pi=93$ MeV) we get the function $m^*=m^*(\Lambda)$ which we use in the $m^*$ equation to get the function $G=G(\Lambda)$. The
result of this analysis is in fig. 1.
\begin{figure}[h]
\begin{center}
\includegraphics[width=5cm]{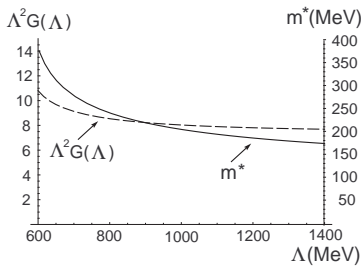}
\end{center}
\caption{\footnotesize The running NJL coupling constant $G$
(dashed line) and the running constituent mass $m^*$ (in MeV,
solid line) as functions of $\Lambda$ (in MeV). $\Lambda$ is the
ultraviolet cutoff. The vertical axis on the left refers to
$\Lambda^2G(\Lambda)$, while the axis on the right refers to
$m^*$. \label{l}}
\end{figure}
Our choice implies  that  a NJL model is defined at any scale by
using the appropriate $G(\Lambda)$. Whereas in the usual case we
have to keep the momenta smaller than the cutoff, now, for any
given momenta, we can fix the cutoff in such a way that it is much
bigger than the  momenta.

\begin{figure}[h]
\begin{center}
\includegraphics[width=5cm]{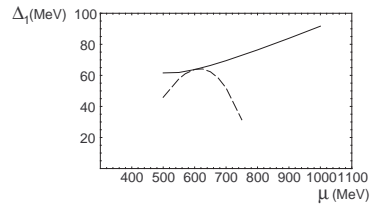}
\end{center}
\caption{\footnotesize The CFL gap $\Delta$ for massless quarks,
as obtained from eq. \ref{2.186}, versus the quark chemical
potential for the two cases discussed in the text. Solid line:
running NJL coupling $G(\mu+\delta)$ and cutoff $\delta=c\mu$,
with $c=0.35$; dashed line: $\delta=\Lambda-\mu$, where
$\Lambda=800$ MeV, and $G(\Lambda)=13.3$ GeV$^{-2}$. The picture
shows the different qualitative behavior with $\mu$ of the gap.
\label{2}}
\end{figure}

In applying these considerations to the calculations at finite
density, we have only to use the appropriate value of the coupling
as given by $G(\mu+\delta)$. To
give an explicit example we consider  the CFL phase with massless
quarks.
 There are  two independent gaps $\Delta$ and $\Delta_9$
($\Delta_9=-2\Delta$ if the pairing is only in the antitriplet
channel) and the gap equations are
\cite{Alford:1997zt,Nardulli:2002ma}: \bea \Delta&=&
-\,\frac{\mu^2G}{6\pi^2} \left(\Delta_9 \ {\rm arcsinh}\frac\delta
{|\Delta_9|}-2\Delta\ {\rm arcsinh}\frac\delta {|\Delta|}
\right)\, ,\cr
 \Delta_9&=& -\,\frac{4\mu^2 G\Delta}{3\pi^2}\
 {\rm arcsinh}\frac\delta
{|\Delta|}\ .\label{2.186}\eea If one uses a fixed value for
$\Lambda=\mu+\delta$, as for instance in ref.
\cite{Steiner:2002gx}, one gets a non monotonic behavior of the
gap, as it can be seen from fig. \ref{2}, (dashed line); a similar
behavior was found in \cite{Steiner:2002gx} (their fig. 1), albeit
a different choice of the parameters produces some numerical
differences.
On the other hand the solid line shows an increasing behavior of
the gap. We see that in this way one reproduces
 qualitatively the behavior found in QCD for asymptotic chemical
 potential \cite{Son:1998uk}.
This result is obtained by the running NJL coupling
$G(\mu+\delta)$, with the following choice of the cutoff $\delta$:
\be \delta=c\,\mu\label{delta}\ee with
 $c$ a fixed constant ($c=0.35$ in fig. \ref{2}). The reason for this choice is that,
as discussed above,  when increasing $\mu$, we do not want to
reduce the ratio of the number of the
  relevant degrees of freedom  to the volume of the Fermi sphere.
Requiring the fractional importance to be constant is equivalent
to require eq. (\ref{delta}).

\section{The CFL phase with massive strange quark}
The massive case is considerably more involved, and the simple set
of equations (\ref{2.186}) has to be substituted by a system of 5
equations \cite{Alford:1999pa} that can be only solved
numerically. In \cite{Steiner:2002gx} the CFL phase with massive
strange quark was also considered. In comparison with
\cite{Alford:1999pa} the derivation we present here has the
advantage of offering semi-analytical results, thanks to an
expansion in powers of $m_s/\mu$. We principally differ from  ref.
\cite{Steiner:2002gx} for the different treatment of the cutoff,
as discussed in the previous section, and for the inclusion of
pairing in both the antitriplet and the sextet color channel. The
possibility of a semi-analytical treatment rests on the HDET
approximation.  This effective lagrangian approach was extended in
\cite{Casalbuoni:2002st} to the 2SC phase with massive quarks and
here we treat  the three flavor case.

Because of limited space we discuss here only the fundamental results; formal details may be found in\cite{Casalbuoni:2003cs}.

\begin{figure}[th]
\begin{center}
\includegraphics[width=5cm]{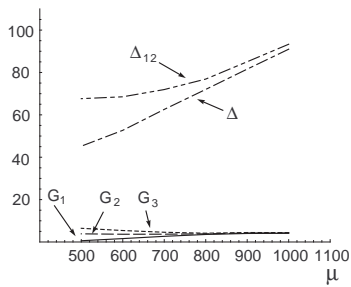}
\end{center}
\caption{\footnotesize The five gap parameters vs. the chemical
potential $\mu$; results obtained by the numerical solution of the
exact gap equations, for $m_s=250$ MeV and $\delta=0.35\mu$. The
upper curve refers to $\Delta_{12}$, while the middle one refers
to $\Delta$. The lower curves are the data obtained for the three
sextet gap parameters $|G_1|$, $|G_2|$ and $|G_3|$. Gaps and $\mu$
are expressed in MeV.} \label{PlotLavoro1}
\end{figure}

We consider the HDET formulation of the theory of CFL condensation in both the antisymmetric $\bar{\bf 3}_A$ and in the symmetric ${\bf
6}_S$ channels. We assume equal masses (actually zero) for the up and down quarks and neglect quark-antiquark chiral condensates, whose
contribution is expected to negligible in the very large $\mu$ limit\footnote{Also the contribution from the repulsive ${\bf 6}_S$ channel
is expected to be small, but we include it because the gap equations
 are consistent only with condensation in both the ${\bf 6}_S$ and the $\bar{\bf
3}_A$ channels.}. The condensate we consider is therefore
\begin{eqnarray}
<0|\psi_{\alpha i}\,C\,\gamma_5\,\,\psi_{\beta j}|0> &\sim& \left(\Delta_{ij}\,\epsilon^{\alpha\beta I}\epsilon_{ijI} \right.\nonumber\\
&+&\left.G_{ij}\,(\delta^{\alpha i}\,\delta^{\beta j}+\delta^{\alpha j}\,\delta^{\beta i}) \right)\ . \label{6plus3condensate}
\end{eqnarray}
The first term on the r.h.s
accounts
 for the condensation in the  $\bar{\bf
3}_A$  channel and the second one describes condensation in ${\bf
6}_S$ channel. As we assume $m_u=m_d$, we put
\begin{eqnarray}
\Delta_{us}=\Delta_{ds}\equiv\Delta\ && \Delta_{ud}\equiv \Delta_{12} \nonumber \\
G_{uu}=G_{du}=G_{ud}=G_{dd}\equiv G_1 && G_{us}=G_{ds}\equiv G_2 \nonumber \\
G_{ss}\equiv G_3 \end{eqnarray}
which reduces the number of independent gap parameters
to five. We stress that we impose electrical and color neutrality,
which, as shown  in ref. \cite{Rajagopal:2000ff},  is indeed
satisfied in the color-flavor locked phase of QCD because in this
phase the three light quarks number densities are equal $
 n_u=n_d=n_s$, with no need for electrons, i.e. $n_e=0$. As a consequence,
the  Fermi momenta of the three quarks are equal: \be p_{F,\,u}\,=\,p_{F,\,d}\,=\,p_{F,\,s}\,\equiv\,p_F\, \ ,\label{p}\ee Armed with the
fermionic lagrangian, one writes the five gap equations whose numerical solutions are reported in Fig.\ref{PlotLavoro1} for $m_s=250$ MeV
and $c=0.35$\cite{Casalbuoni:2003cs}.

One notes that the effects of the strange quark mass are of order $m_s^2/\mu^2$, so a semi-analytical solution can be found by performing
an expansion in the strange quark mass $x_s=\frac{m_s}{\mu_s} \approx\frac{m_s}{\mu}\ll 1$. Here $\mu=\mu_u=\mu_d$, $\mu_s=\mu+\delta\mu$
with $\delta\mu={\cal O}(m_s/\mu)^2$. As shown in Ref.\cite{Casalbuoni:2003cs}, in this perturbative frame the set of five gap equations
reduces to a set of five {\em algebraic} equations plus the two trascendental gap equations for the massless case. In
Ref.\cite{Casalbuoni:2003cs} we find that the agreement between the exact and the approximated solutions of the gap equations is indeed
good.

\section{The 2SC phase with massive flavors}
Briefly we describe some results obtained within the formalism of HDET for the 2SC phase of QCD with $u$ and $d$ massive quarks.
Such a phase is characterized by the condensation pattern
\begin{equation}
<0|\psi_{\alpha i}\,C\,\gamma_5\,\,\psi_{\beta j}|0>\sim
\Delta\,\epsilon^{\alpha\beta 3}\epsilon_{ij3}
\label{2SCcond}
\end{equation}
We work in the hypothesis $p_{F,u}=\mu_u\,v_u=p_{F,d}=\mu_d\,v_d$, which implies a difference in the chemical potentials of the
two flavors when their masses are unequals. In this case one has only one gap parameter, and the solution of the gap equation may be written
explicitely\cite{Casalbuoni:2002st}.

We wish to discuss here some property of the $U(1)_A$ pseudo-Goldstone mode of the 2SC phase. It is well known that the condensation pattern
in Eq.\eqref{2SCcond} breaks spontaneosly  $U(1)_A$, which is however also broken by the strong anomaly. At high $\mu$ the latter breaking is soft
and one expects that the associated pseudo-Goldstone, called $\eta'$, is almost massless.
In HDET one gets the dynamical properties of the pseudo-Goldstone by bosonization of fermionic lines\cite{Casalbuoni:2002st}.
This results in a one-loop effective lagrangian.
As noticed in\cite{Casalbuoni:2002st},\cite{Beane:2000ms} and\cite{Schafer:2001za} the mass term of the $\eta'$ is due to antigap insertions,
while the gap insertions give rise to the kinetic term.
Within the HDET one obtains the mass formula (see also\cite{Beane:2000ms},\cite{Schafer:2001za}):
\begin{equation}
m^2_{\eta'}=4\,\Delta^2\,\frac{m_u m_d}{\mu^2}\log\left(\frac{\mu}{\Delta}\right)
\label{massETA}
\end{equation}
To this result one should add the contribution of the instantons. However, the latter has been already
estimated\cite{Son:2001ss},\cite{Son:2001jm},\cite{Son:2000fh} and the result is that at asymptotic $\mu$ the instanton contribution is
negligible with respect to the quark massive one (other interesting studies of the role of instantons in high density or large $N_c$ QCD
may be found in\cite{Schafer:2002ty}).

Moreover, we get formulas for the $\eta'$ velocity and its decay constant
\begin{equation}
v^2_{\eta'}=\frac{|v_u||v_d|}{3}~~~~~~~~~~~~~
f_{\eta'}^2=\frac{8\mu_u \mu_d}{\pi^2}\frac{|v_u||v_d|}{|v_u|+|v_d|}
\label{dyna}
\end{equation}
In particular, the $v_{eta'}^2$ should be compared with the massless value $v^2_{\eta'}=1/3$\cite{Nardulli:2002ma}. The last equations are
new results, which go beyond the expansion in $m/\Delta$ commonly used in the computation of the mass effects in CSC.

\section{Conclusions and outlooks}
In this paper we proposed a modified NJL model with running coupling constant, which allows to overcome the troubles of the zero density fixed cutoff scheme. Moreover,
we have discussed several points relative to massive quarks effects in two CSC phases of QCD, in the framework of HDET.

For the CFL phase with a massive strange quark we have numerically solved the gap equations for condensation both in the triplet and in the
sextet channels. Such a solutions explicitely show that the massive effects are of order $x_s^2=m_s^2/\mu^2$; as a consequence, a
perturbative expansion in terms of $x_s^2$ is meaningful. The perturbative gap equations are a set of algebraic equations, as opposite to
the full gap equations which are trascendental. We show in Ref.\cite{Casalbuoni:2003cs} that there exists a good agreement between the
exact and the perturbative gap equations. This perturbative approach could be applied for example to get analytic results on the massive
states of CFL, as for example gluons (in a recent paper\cite{Jackson:2003dk} a non perturbative study of the massive spectrum of the CFL
phase is presented).

Turning to the 2SC phase, we have solved the gap equation for both massive flavors: in HDET one gets a closed formula which relates the gap parameter to the quark masses.

Lastly but not least important, we shown HDET results for the dynamical properties of the pseudo-Goldstone $\eta'$, obtaining
a leading order mass formula and expressions for velocity and decay constant which go beyond the usual $m/\Delta$ expansion
used in the massive effects in the CSC calculations.

\section*{Acknowledgments}
We wish to thank R. Casalbuoni, F. De Fazio, R. Gatto and G. Nardulli. Moreover, it is a pleasure to thank
M. Mannarelli, F. Sannino and T. Schafer for useful discussions and very interesting observations about some points of the work presented in this paper.

\end{document}